\documentstyle[12pt]{article}
\def\lapprox{\lower .7ex\hbox{$\;\stackrel{\textstyle <}{\sim}\;$}}
\def\half{\frac{1}{2}}
\def\d{{\rm d}}

\begin{document}
\begin{titlepage}
\vspace*{-1cm}
\begin{flushright}
DTP/95/82   \\
December 1995 \\
\end{flushright}
\vskip 1.cm
\begin{center}
{\Large\bf Polarized Parton Distributions in the Nucleon}
\vskip 1.cm
{\large  T.~Gehrmann and W.J.~Stirling}
\vskip .4cm
{\it Departments of Physics and Mathematical Sciences, University of Durham \\
Durham DH1 3LE, England }\\
\vskip 1cm
\end{center}

\begin{abstract}
The distribution of the spin of the nucleon among its constituents
can be paramet\-rized in the form of polarized parton distribution functions
 for quarks and gluons. Using all available data on the polarized structure
function $g_1(x,Q^2)$, we determine these distributions both at leading and
next--to--leading order in perturbation theory.
 We suggest three different, equally possible
scenarios for the polarized gluon distribution, which is found to be
only loosely constrained by current  experimental data.
We examine various possibilities of measuring polarized parton
distributions at future experiments.
\end{abstract}
\vfill
\end{titlepage}
\newpage

\section{Introduction}

Understanding the spin structure of the nucleon is one of the most
important questions in strong interaction physics today. There has been
renewed interest in recent years, largely as a result of a series of
high--precision  polarized deep inelastic scattering experiments.
So far, the bulk of the information comes from measurements of the
spin-dependent structure function $g_1(x,Q^2)$.
 From such measurements, spin-dependent
 parton distributions can be extracted and compared with theoretical
 models, or used to make predictions for future experiments.
Several theoretical issues have received particular attention: the
predictions of various sum--rules and the measurement of $\alpha_s$,
the magnitude of the polarized gluon
distribution, and the behaviour of the polarized parton distributions
at small $x$.

In a previous study \cite{gs} we performed a global analysis of all available
deep inelastic polarized structure function
data in the context of the QCD--improved parton model at
leading order. We supplemented the experimental information with
theoretical assumptions about the flavour content and the form of
the distributions  at large and small $x$. In particular, we presented
three different sets of parton distributions characterized by qualitiatively
different polarized gluon distributions, and suggested further experiments
which could discriminate between them.

Since the work of Ref.~\cite{gs} there have been several important advances.
More precise polarized structure function data has become available.
Both the the SMC \cite{SMC_d} and SLAC--E143 collaborations \cite{SLAC-e143}
 have measured the structure function
$g_1$ of the deuteron,  supplementing their previous proton target
measurements, and the SLAC--E143 collaboration has
also measured the structure function $g_2$ on proton and deuteron targets
\cite{SLAC-e143_g2}. The new $g_1$ data are particularly important for
determining the  polarized $u$-- and $d$--quark distributions separately.

A second advance has been the theoretical calculation \cite{vn2} of
the next--to--leading order contributions to the polarized splitting
functions $\Delta P_{ij}(x,\alpha_s)$ which determine the $Q^2$ evolution
of the distributions. This not only improves the precision of the
phenomenology, but also allows for the first time a consistent
factorization/renormalization scheme dependence to be imposed on the analysis.

In this study we repeat the analysis of \cite{gs} using the  new polarized
structure function data, but now at both leading and next--to--leading
order in perturbation theory.
Our aim, as before, is to derive a consistent set of polarized
parton distributions. At the same time we can check that the $Q^2$ dependence
seen in the data is consistent with the predictions of perturbative QCD.
We discuss in detail the constraining power of the various data sets, the role
of the polarized gluon distribution, and the prospects for future measurements.
A similar analysis of polarized
structure functions has been reported recently in
Ref.~\cite{GRSV}, using the `dynamical parton model'
approach in which the distributions at small $x$ are
generated dynamically from valence--like distributions
at a small starting $Q_0^2$ scale. A reasonable
description of the data is obtained. Our approach,
in contrast, is to constrain the
small $x$ distributions from the data, without prejudice
as to the origin of the observed behaviour. This is in the same
spirit as the unpolarized analysis of Refs.~\cite{mrsg,cteq}.

We begin by reviewing the basic theoretical input to the analysis.
The fundamental quantity of interest is the structure function $g_1$ which, in
analogy with the unpolarized structure function $F_1$, can (in the
QCD--improved parton model at leading order)
be expressed in terms of the probability distributions for finding quarks
 with spin parallel or
antiparallel to the longitudinally polarized parent nucleon:
\begin{eqnarray}
F_1(x,Q^2) &=& \half \sum_q\; e_q^2\;  [q(x,Q^2) +\bar q(x,Q^2) ]  \\
g_1(x,Q^2) &=& \half \sum_q\; e_q^2\;  [\Delta q(x,Q^2) +\Delta\bar q(x,Q^2) ]
,
\label{naiveg1}
\end{eqnarray}
where
\def\qup{q_{\uparrow}}
\def\qdown{q_{\downarrow}}
\begin{equation}
q = \qup + \qdown\; , \quad \Delta q = \qup-\qdown \; .
\end{equation}

The first moment of $g_1(x,Q^2)$ measures the expectation value of the
axial--vector current between two identical nucleon states:
\begin{equation}
\Gamma_1^{p,n} = \pm \frac{1}{12} a_3 + \frac{1}{36} a_8 + \frac{1}{9} a_0,
\end{equation}
where the conserved nonsinglet axial vector currents
are known from the
hyperon decay constants (the Ellis--Jaffe sum rule \cite{gour,ej}):
\begin{eqnarray}
a_3 & = & F+D \nonumber\\
a_8 & = & 3F-D.
\end{eqnarray}
In contrast, the singlet axial vector current $a_0$ is non--conserved due to
the axial anomaly. It cannot be related to any other experimental
quantity, although several model--dependent estimates of it can be found in the
literature: if the total nucleon spin is carried only by quarks one
finds $a_0=1$ \cite{gour}, and if the strange quark sea in the nucleon is
unpolarized then $a_0=a_8\approx 0.579$ \cite{ej}. Both these
estimates are clearly
ruled out by  experimental measurements \cite{SMC_d,SLAC-e143,exp}, which find
$a_0 \approx
0.15 - 0.3$
in the range $3\;\mbox{GeV}^2 < Q^2 < 12 \;\mbox{GeV}^2$.

In the QCD--improved parton model $g_1$ is expressed in terms
of  parton distributions
for the polarization of quarks and gluons,
\begin{eqnarray}
g_1(x,Q^2) &=& {1\over 2}\; \sum_q\; e_q^2 \; \int_x^1 {dy\over y}
\;  \left[\Delta q(x/y, Q^2) +\Delta\bar q(x/y,Q^2) \right]\; \nonumber \\
& & \times  \left\{\delta(1-y)
+ {\alpha_s(Q^2) \over 2\pi} \Delta C_q(y) + \ldots \right\} \nonumber \\
& & + \; \frac{\langle e_q^2 \rangle}{2} \int_x^1 \; {dy\over y}\;  \Delta
G(x/y,Q^2) \; \left\{
n_f\;   {\alpha_s(Q^2) \over 2\pi}\;\Delta C_G(y) +\ldots \right\}.
\label{qcdg1}
\end{eqnarray}

The variation of these polarized parton distributions with
$Q^2$ is determined by the GLAP evolution equation \cite{ap}:
\begin{eqnarray}
\label{evo}
\lefteqn{\frac{\partial}{\partial \ln Q^2}
\left(\begin{array}{c} \Delta q \\ \Delta G
\end{array} \right) (x,Q^2)} \nonumber \\
& & =
\frac{\alpha_s(Q^2)}{2\pi}
\int_x^1 \frac{\d y}{y} \left(\begin{array}{cc} \Delta P_{qq}  & \Delta
P_{qG} \\ \sum_q \Delta P_{Gq} & \Delta P_{GG}\end{array} \right) (y)\;
\left(\begin{array}{c} \Delta
q \\ \Delta G \end{array} \right)(x/y,Q^2).
\end{eqnarray}

For a  consistent theoretical  description of $g_1(x,Q^2)$,  the coefficient
functions in (\ref{qcdg1}) must be truncated at the same order in
perturbation theory  as the
splitting functions in (\ref{evo}). Although the ${\cal O}(\alpha_s)$
corrections to the coefficient functions $\Delta C_q(y)$ \cite{cqnlo} and
$\Delta C_G(y)$ \cite{cgnlo} have been known for a long time, a
next--to--leading order study was not possible until very recently, as
the ${\cal O}(\alpha_s)$ corrections to the $\Delta P_{ij}(y)$
were not known. These have now
been published for the first time in Ref.~\cite{vn2}.

Using the next--to--leading order evolution equations \cite{evonlo}, the
{\it unpolarized} parton distributions can be determined to a
high level of accuracy
(see for example Refs.~\cite{mrsg,cteq,grv94}) using a wide variety
 of high--precision experimental data
from  different processes. As the only data available on the
spin structure of the nucleon are the structure function measurements
of Refs.~\cite{SMC_d,SLAC-e143,exp}, we are still far from obtaining a
similar level of precision  on the polarized distributions.

The appropriate choice of scheme for $g_1$ in the
QCD--improved parton model has been a matter of discussion over the
last few years (see for example \cite{physrep} for a recent review). The main
issue is the treatment of the gluonic contributions to the Ellis--Jaffe
sum rule. While the nonsinglet contributions to this sum rule can be
related to conserved currents and are unambiguously defined, there is an
arbitrariness in the decomposition of the singlet contribution
\cite{cgnlo,lamli}, depending on the absorption procedure used for the
axial anomaly. In the $\overline{{\rm MS}}$ scheme, this anomaly is
absorbed into the next--to--leading order pure singlet quark--to--quark
splitting function \cite{vn1}. Another possible procedure is to
allow for an anomalous gluonic contribution to the Ellis--Jaffe sum
rule, which is absent in the  $\overline{{\rm MS}}$-scheme. The
appropriate scheme transformation \cite{vn1} then yields conservation
of the quark contribution to $a_0$. In the present study
we will work entirely in the $\overline{{\rm MS}}$ scheme. Any other
choice of scheme would tie the resulting distributions to $g_1$, as
higher--order corrections to the coefficient functions of other
processes are so far only known in the
$\overline{{\rm MS}}$ scheme.\footnote{A NLO analysis of polarized deep
inelastic scattering data in the so-called Adler--Bardeen scheme, which
displays different factorization properties, has recently  been carried out
\protect\cite{BFR95}. In  Ref.~\protect\cite{BFR95} the emphasis
is on the determination
of the singlet axial charge $a_0$ and the first moment of the polarized
gluon distribution.}

The choice of the $\overline{{\rm MS}}$ scheme also allows us to use
our distributions in conjunction with a broad range of unpolarized
distributions, the majority of which are defined in this scheme.
For reference, the $\overline{{\rm MS}}$ scheme coefficient functions
in (\ref{qcdg1}) are
\begin{eqnarray}
\Delta C_q (y)  & = & C_F \left\{ 2 \left(\frac{\ln (1-y)}{1-y}\right)_{+}
             - \frac{3}{2} \right.
              \left(\frac{1}{1-y}\right)_{+} -(1+y)\ln (1-y) \nonumber \\
& & \left. -\frac{1+y^2}{1-y} \ln y + 2 + y +
        \delta (1-y) (-2 \zeta (2)-\frac{9}{2}) \right\}
              \nonumber\\
\Delta C_G (y) & = & T_R \left\{ 2 (2y-1) \ln \left( \frac{1-y}{y}
                    \right) + 2 (3-4y) \right\}
\end{eqnarray}

In the following section we will present the results of our leading
and next--to--leading order parton model analyses of polarized
structure function data. Section~3 discusses the prospects of various
future experiments on polarized nucleons and Section~4 contains a
brief summary and our conclusions.

\section{Polarized parton distributions in leading and next--to--leading
order}

We adopt a similar approach to the global analysis  of unpolarized
parton distributions in the nucleon \cite{mrsg} by parametrizing the polarized
distributions at the starting scale in the form:
\begin{eqnarray}
x\Delta u_v (x,Q_0^2) & = & \eta_u A_u x^{a_u} (1-x)^{b_u}
(1+\gamma_u x + \rho_u x^{1/2}) \nonumber \\
x\Delta d_v (x,Q_0^2) & = & \eta_d A_d x^{a_d} (1-x)^{b_d}
(1+\gamma_d x + \rho_d x^{1/2}) \nonumber \\
x\Delta \bar{q} (x,Q_0^2) & = & \eta_{\bar q} A_{\bar q}
 x^{a_{\bar q}} (1-x)^{b_{\bar q}}
(1+\gamma_{\bar q} x + \rho_{\bar q} x^{1/2}) \nonumber \\
x\Delta G (x,Q_0^2) & = & \eta_G A_G x^{a_G} (1-x)^{b_G}
(1+\gamma_G x + \rho_G x^{1/2}),
\label{pform}
\end{eqnarray}
where we take $Q_0^2 = 4$~GeV$^2$. The normalization factors are
\begin{equation}
A_f^{-1} (f=q,G) = \left(1+\gamma_f\frac{a_f}{a_f+b_f+1}\right)
\frac{\Gamma(a_f)\Gamma(b_f+1)}{\Gamma(a_f+b_f+1)}+ \rho_f
\frac{\Gamma(a_f+0.5)\Gamma(b_f+1)}{\Gamma(a_f+b_f+1.5)}
\end{equation}
which ensures that the first moments of the distributions, $\int_0^1dx
\Delta f(x,Q_0^2)$, are given by $\eta_f$.

Various experimental measurements of unpolarized lepton($=e,\mu,\nu$)--nucleon
and unpolarized Drell--Yan cross sections yield a reasonably
precise flavour decomposition of the light quark $(u,d,s)$ sea.
Such a decomposition is not yet possible for the polarization of the light
quark sea, as measurements of the structure function $g_1$ are only
sensitive to the charge weighted sum of all quark flavours, not to the
individual distributions. We therefore
assume a SU(3)--symmetric antiquark polarization
$\Delta \bar{q} (x,Q_0^2)= \Delta \bar{u} (x,Q_0^2)= \Delta \bar{d}
(x,Q_0^2)= \Delta \bar{s} (x,Q_0^2)$. This {\it ad--hoc} assumption is only
justified at the present level of experimental knowledge, and is
furthermore immediately broken by next--to--leading order evolution
\cite{ross}.

The first moments of the polarized quark distributions can be
determined from the measured values of the Ellis--Jaffe sum rule.
Imposing SU(3)--symmetry at $Q_0^2$, this sum rule reads
\begin{equation}
\label{eq:gamma1}
\Gamma^{p,n}_1 (Q_0^2) = \left(1-\frac{\alpha_s(Q_0^2)}{\pi}\right)
\left(\pm
\frac{1}{12} a_3 + \frac{1}{36} a_8 + \frac{1}{9} a_0 (Q_0^2) \right)
\end{equation}
with
\begin{eqnarray}
a_3 & = & \eta_u - \eta_d = F+D , \nonumber \\
a_8 & = & \eta_u + \eta_d = 3F-D , \nonumber \\
a_0 (Q_0^2) & = & \eta_u + \eta_d + 6 \eta_{\bar{q}} (Q_0^2) .
\end{eqnarray}
In this approach,  the first moments of the valence quark
polarizations are obtained from the nonsinglet axial--vector current matrix
elements \cite{clo93}, while the first moment of the sea quark distribution is
inferred from the measured value of $\Gamma_1$. For the
leading--order (LO) distributions, we correct the normalization of $\eta_u$
and $\eta_d$ by the ${\cal O} (\alpha_s)$ coefficient function  in
(\ref{eq:gamma1}) \cite{gs}. The first moments
obtained by this procedure are listed in Table~\ref{tab:fixed}.  Note
 that the Ellis--Jaffe sum rule is a ($Q^2$--independent) constant at leading
order in perturbation theory, as the leading-order coefficient functions
are only expanded up to ${\cal O} (\alpha_s^0)$
and scaling violations in $\eta_{\bar{q}}(Q^2)$ arise only from
the splitting functions at next--to--leading order.

The polarized gluon distribution  enters $g_1(x,Q^2)$ at next--to--leading
order. It is only very weakly constrained so far, as no
experimental data are available on gluon--initiated processes such as
direct--$\gamma$ or heavy meson production.
The polarized gluon distribution $\Delta G(x,Q^2)$ is therefore
not well--determined by a fit to the $g_1$ data alone, and so additional
theoretical constraints have to be applied.
 It follows from the structure of the
polarized splitting functions that the small-$x$ behaviour of the
gluon and sea quark distributions are closely correlated, which
justifies the assumption $a_G=a_{\bar q}$ in (\ref{pform}).
 In the region $x>0.1$, structure
functions and their evolution
are dominated by valence quark contributions, and the impact of the gluon
is completely negligible. In Ref.~\cite{gs} we explored various possibilities
for the form of $\Delta G(x,Q^2)$ at large $x$: hard and soft distributions
with the
spin aligned with that of the parent hadron, and a distribution with the spin
anti--aligned. All three choices give equally good descriptions of the
structure function data, but would be relatively easy to discriminate
if data on polarized gluon--initiated processes were available.
We adopt the same procedure here, i.e. we consider three
 equally possible scenarios for the behaviour of $\Delta
G(x,Q_0^2)$, which can be parametrized as follows:
\begin{eqnarray}
\mbox{Gluon A}\; : & & \gamma_G=0,\ \  \quad \rho_G=0,\nonumber\\
\mbox{Gluon B}\; : & & \gamma_G=-1,\quad \rho_G=2,\nonumber\\
\mbox{Gluon C}\; : & & \gamma_G=0,\ \  \quad \rho_G=-3.
\end{eqnarray}
Due to the introduction of the additional parameter $\rho_G$
in the starting parametrizations (\ref{pform}), these
distributions look slightly different from the ones presented in \cite{gs}.

The normalization $\eta_G$ of the gluon distribution can only be
determined consistently from the
experimental data in a next--to--leading order analysis,
 where it still has
a large error.
At leading order, we can estimate $\eta_G$ by attributing {\it all}
the violation of
the Ellis--Jaffe sum rule to a large gluon polarization and vanishing
sea quark polarization. In this way we obtain $\eta_G=1.9$,
 only slightly different
from the value 1.971 obtained in Ref.~\cite{gs}.
Note, however, that at leading order the apportioning of the singlet
contribution to $\Gamma_1$ between gluons and sea quarks is completely
arbitrary \cite{vn1}. In fact we shall see below that a consistent NLO
treatment
gives a value of $\eta_G$ somewhat less than our estimated leading
order value, and similar to the range of values found in Ref.~\cite{BFR95}.

If parton distributions are interpreted in the probabilistic picture of
the naive parton model, the magnitude of the polarized distributions
cannot exceed the unpolarized distributions, in order to guarantee
positive probabilities for the individual polarization
states, i.e.
\begin{equation}
\label{pos}
\vert \Delta f(x) \vert \; < \; f(x), \qquad (f=q,G).
\end{equation}
This is in fact only a rigid
constraint at leading order, since parton distibutions at higher
orders are only scheme--dependent constants of renormalization and not
strict probability densities. The fundamental constraint at arbitrary orders
in perturbation theory is the
positivity of physical cross sections for all possible helicity
configurations, which does not necessarily imply the positivity of the
distributions.

Positivity of the polarized distributions is achieved by constraining
the parameters of the starting distributions at $Q_0^2$ such that
\begin{equation}
\vert \Delta f(x,Q_0^2) \vert \; < \;  f(x,Q_0^2), \qquad (f=q,G).
\end{equation}
Perturbative evolution preserves the positivity of the individual
helicity distributions, hence (\ref{pos}) is fulfilled at any $Q^2$.

In our leading--order analysis, we use the unpolarized distributions from
\cite{grv94} for reference. At $Q_0^2=4 \;\mbox{GeV}^2$ these are
\begin{eqnarray}
xu_{v} (x,Q_0^2) & = & 3.221 \; x^{0.564} (1-x)^{3.726}
(1-0.6889x^{0.200}+2.254x+1.261x^{3/2}) \nonumber \\
xd_{v} (x,Q_0^2) & = & 0.507 \; x^{0.376} (1-x)^{4.476}
(1+1.615x^{0.553}+3.651x+1.299x^{3/2}) \nonumber \\
x(\bar{u}+\bar{d}) (x,Q_0^2) & = & \Big[x^{0.158}(0.738-0.981x+1.063x^2)
(-\ln x)^{0.037} + \nonumber\\
& & 0.00285 \exp \left(\sqrt{-4.010\ln x}\right) \Big]
(1-x)^{6.356} \nonumber\\
xs (x,Q_0^2) & = & 0.0034 (-\ln x)^{-1.15} (1-2.392x^{1/2}+7.094x)
(1-x)^{6.166} \nonumber \\
& & \exp \left(\sqrt{-6.719\ln x}\right) \nonumber \\
xG (x,Q_0^2) & = & \Big[ x^{0.731}(5.110-1.204x-1.911x^2)
(-\ln x)^{-0.4718} + \nonumber\\
& & 0.0527 \exp \left(\sqrt{-4.584\ln x}\right) \Big]
(1-x)^{5.566}
\end{eqnarray}
The reference unpolarized distributions at next-to-leading order
are  the
A$'$ set of \cite{mrsg}, which are parametrized at $Q_0^2=4
\;\mbox{GeV}^2$ as
\begin{eqnarray}
xu_{v} (x,Q_0^2) & = & 2.26 x^{0.559} \; (1-x)^{3.96} \;
(1-0.54x^{1/2}+4.65x) \nonumber\\
xd_{v} (x,Q_0^2) & = & 0.279 x^{0.335} \; (1-x)^{4.46} \;
(1+6.80x^{1/2}+1.93x) \nonumber\\
x\mbox{Sea}(x,Q_0^2) & = & 0.956 x^{-0.17}  \; (1-x)^{9.63} \;
(1-2.55x^{1/2}+11.2x) \nonumber\\
xG(x,Q_0^2) & = & 1.94 x^{-0.17}  \; (1-x)^{5.33} \;
(1-1.90x^{1/2}+4.07x)
\end{eqnarray}
Note that the choice of unpolarized distributions is not crucial
for the present analysis. All the widely available leading and
next--to--leading
order distributions provide very good fits to the unpolarized structure
function
data, and the small differences between them are much smaller than the
precision with which the polarized distributions are currently determined.
The flavour decomposition of the unpolarized sea quark distributions is
also unimportant for our present analysis and will therefore be
disregarded. The starting  sea quark distribution of Ref.~\cite{mrsg}
contains a very small  charm quark contribution which can safely be
ignored in the present analysis.
To be consistent with the evolution of the unpolarized distributions,
we take
\begin{equation}
\Lambda_{\rm LO}^{n_f=4} = 200 \;\mbox{MeV}\ \cite{grv94}, \qquad
\Lambda_{\rm NLO}^{n_f=4} = 231 \;\mbox{MeV}\ \cite{mrsg}.
\end{equation}
These correspond to $\alpha_s(M_Z^2)=0.123$ (LO) and $\alpha_s(M_Z^2)=0.112$
(NLO).

The parameters most affected by the positivity constraints are the
large-$x$ exponents $b_f$. For the valence quarks, we fix
$b_u=b_u(\mbox{unpol.})$ and $b_d=b_u(\mbox{unpol.})+1$, motivated
by counting rule estimates \cite{bro}. The parameters $b_G$ and $b_{\bar q}$
are constrained to be at least as large as their unpolarized
counterparts in the fit, but this constraint has only minimal impact.
We find that only the $\Delta
d_{v}(x)$ distribution tends to saturate positivity, requiring
the combination $\gamma_d+\rho_d$ to be limited in the fit.

The data currently available on $g_1$  are not able
to test the various theoretical model predictions for the  small-$x$
behaviour of the polarized parton distributions~\cite{lowxth},
which  are only expected
to apply  at much lower values of $x$ \cite{lowx}.
The parameters $a_f$ are therefore
only effective exponents valid over some finite interval in $x$.
It therefore makes no sense to
postulate positivity for $x\rightarrow 0$ by constraining the $a_f$.

\begin{table}[htb]
\begin{center}
\begin{tabular}{|c|r|r|} \hline
\rule[-1.2ex]{0mm}{4ex}&  LO  & NLO \\ \hline
\rule[-1.2ex]{0mm}{4ex}$\eta_u$ & 0.823 & 0.918 \\ \hline
\rule[-1.2ex]{0mm}{4ex}$\eta_d$ & $-$0.303 & $-$0.339 \\ \hline
\rule[-1.2ex]{0mm}{4ex}$\eta_G$ & 1.9 &  \\ \hline
\rule[-1.2ex]{0mm}{4ex}$\eta_{\bar{q}}$ & $-$0.0495 & $-$0.060 \\ \hline
\rule[-1.2ex]{0mm}{4ex}$b_u$ & 3.73 & 3.96 \\ \hline
\rule[-1.2ex]{0mm}{4ex}$b_d$ & 4.73 & 4.96 \\ \hline
\end{tabular}
\caption{Fixed parameters in the LO and NLO($\overline{{\rm MS}}$)
polarized parton distribution fits.}
\label{tab:fixed}
\end{center}
\end{table}

The contribution of charmed quarks to $g_1(x,Q^2)$ is negligible at
present experimental energies \cite{grvchm} and will not be considered
in this analysis. We therefore adopt the evolution procedure of
Ref.~\cite{grv94} and fix the number of flavours in the splitting functions
at $n_f=3$, while the number of flavours in $\alpha_s$ increases at
each mass threshold,
\begin{equation}
m_c = 1.5 \;\mbox{GeV}, \qquad m_b=4.5 \;\mbox{GeV}, \qquad m_t=180
\;\mbox{GeV},
\end{equation}
and $\Lambda(n_f)$ is determined by requiring $\alpha_s$
to be continuous across each threshold.

Rather than measuring $g_1(x,Q^2)$ directly from absolute cross section
differences, it is  the relative asymmetry
\begin{equation}
A_1 (x,Q^2) \simeq \frac{g_1(x,Q^2)}{F_1(x,Q^2)}
\label{a1def}
\end{equation}
which is determined experimentally. The structure function
$g_1(x,Q^2)$ is then inferred using a
particular parametrization of $F_1(x,Q^2)$. Some
experimental groups assume $Q^2$--scaling of $A_1(x,Q^2)$ in their
extraction of $g_1(x,Q^2)$. In order to
have a consistent set of data, we have
instead  used the measured values of $A_1(x,Q^2)$ quoted by the
experiments and re-evaluated $g_1(x,Q^2)$ from Eq.~(\ref{a1def}),
constructing $F_1(x,Q^2)$
from the parametrizations of $F_2$ \cite{nmcf2} and $R$ \cite{slacr}
which were used in the most recent measurements.

Applying the constraints outlined above, we have used all available
world data on $A_1^{p,d,n}(x,Q^2)$ \cite{SMC_d,SLAC-e143,exp}
to fit the polarized quark and gluon distributions with the parametric
forms of (\ref{pform}) imposed at
$Q_0^2= 4\;\mbox{GeV}^2$. About 35\% of these data were taken at
$Q^2<Q_0^2$. To have sensible constraints on the distributions, in
particular for $x<0.02$, we  include these datapoints
in the global fit.
The distributions in the region $1\; \mbox{GeV}^2 < Q^2 < Q_0^2$ are
obtained by inverting the evolution matrix, which is straightforward in
$N$-moment space.

A problem with using low $Q^2$ data points in the fit is the
possible contamination by higher--twist contributions.
We have tried to estimate to magnitude of such contributions to
$g_1$ using the parametrization of
$F_2^{\rm HT}$ from Ref.~\cite{bcdms} and assuming
$g_1(x,Q^2)^{\rm HT}\approx
g_1(x,Q^2)^{\rm LT} (1+C^{\rm HT}(x)/Q^2)$. The higher--twist
contributions estimated in this way are found to be small
for all data--points apart from the two lowest $x$  bins of the SMC
experiment.

The global fit is performed using GLAP evolution algorithms
in  $N$-moment space \cite{ourevo}. The distributions and
structure functions are then restored by a numerical inversion into
$x$ space.
The results of the global fit using the leading and next--to--leading
order $(\overline{{\rm MS}})$
expressions for the splitting functions and the $g_1$ coefficient
functions are listed in Table~\ref{tab:res}. The resulting distributions at
$Q_0^2$ are shown in Figs.~1 (LO) and 2 (NLO).

\begin{table}[htb]
\begin{center}
\begin{tabular}{|c|r|r|r|r|r|r|} \hline
\rule[-1.2ex]{0mm}{4ex}&  A (LO)  & B (LO) & C (LO) & A (NLO) &
B(NLO) & C (NLO) \\ \hline
\rule[-1.2ex]{0mm}{4ex}$a_u$ & 0.578 & 0.585 & 0.582 & 0.512 & 0.504 &
0.471 \\ \hline
\rule[-1.2ex]{0mm}{4ex}$\gamma_u$ & 9.38 & 9.31 & 9.50 & 11.65 & 11.98
& 13.14 \\ \hline
\rule[-1.2ex]{0mm}{4ex}$\rho_u$ & $-$4.26 & $-$4.28 & $-$4.28 & $-$4.60 &
$-$4.61 & $-$4.90 \\ \hline
\rule[-1.2ex]{0mm}{4ex}$a_d$ & 0.666 & 0.662 & 0.660 & 0.780 & 0.777 &
0.809 \\ \hline
\rule[-1.2ex]{0mm}{4ex}$\gamma_d$ & 10.46 & 10.91 & 11.04 & 7.81 &
8.18 & 6.73 \\ \hline
\rule[-1.2ex]{0mm}{4ex}$\rho_d$ & $-$5.10 & $-$5.09 & $-$5.06 & $-$3.48 &
$-$3.61 & $-$1.99  \\ \hline
\rule[-1.2ex]{0mm}{4ex}$\eta_G$ & & & & 1.71 & 1.63 & 1.02  \\ \hline
\rule[-1.2ex]{0mm}{4ex}$a_G$ & 0.520 & 0.524 & 0.456 & 0.724 & 0.670 &
0.425 \\ \hline
\rule[-1.2ex]{0mm}{4ex}$b_G$ & 9.45 & 6.87 & 8.72 & 5.71 & 5.34 &
11.05 \\ \hline
\rule[-1.2ex]{0mm}{4ex}$b_{\bar{q}}$ & 15.06 & 15.96 & 11.82 & 14.40 &
18.06 & 16.40 \\ \hline
\rule[-1.2ex]{0mm}{4ex}$\gamma_{\bar{q}}$ & 2.30 & 2.42 & 2.11 & 4.63
& 5.30 & $-$2.67 \\ \hline
\rule[-1.2ex]{0mm}{4ex}$\rho_{\bar{q}}$ & $-$2.00 & $-$2.00 & $-$1.95 &
$-$4.96 & $-$5.25 & $-$3.08 \\ \hline \hline
\rule[-1.2ex]{0mm}{4ex}$\chi^2$ & 98.3 & 97.7 & 100.0 & 89.7 & 91.0 &
93.4 \\ \hline
\end{tabular}
\caption{Fitted parameters in the LO and NLO($\overline{{\rm MS}}$)
polarized parton distributions at $Q_0^2$.
The $\chi^2$ values are with respect
to the 110 data points included in the global fit.}
\label{tab:res}
\end{center}
\end{table}

The resulting parameters are not independent of each other.
In particular,    $a_u$,
$a_d$ and $a_G=a_{\bar{q}}$ are strongly correlated. The $a_i$ of the valence
 distributions are anticorrelated with the corresponding $\gamma_i$,
reflecting the fact that $a_i$ is only an effective exponent for
a finite range in $x$. The $\gamma_i$ and $\rho_i$ are also
anticorrelated.
The $\chi^2$ distribution is very flat  around the local
minima found by the global fits, especially with respect to
 the gluon and sea quark parameters.

The three gluon scenarios give fits of almost identical quality,
reflecting the small impact of the gluon distribution on $g_1(x,Q^2)$
at large and medium $x$. The $\chi^2$ obtained in the NLO fits are
systematically lower due to the additional degree of freedom given by
the normalization of the polarized gluon distribution.
All fits give very  good descriptions for the polarized structure functions
$g_1^{p,n,d} (x,Q^2)$. This is illustrated in Fig.~3, which shows the NLO
description of the various $g_1$ measurements using Gluon A.
The curves correspond to $Q^2 = 1,4,10,50$~GeV$^2$, reflecting the
spread in $Q^2$ values of the different data sets. There is a systematic
decrease in the $Q^2$ values of the data points from large $x$ to small
$x$.

The contributions of $u_v(x,Q^2)$ and $d_v(x,Q^2)$ to the
neutron structure function $g_1^n(x,Q^2)$ are almost equal in
magnitude but opposite in sign. The neutron structure function is
therefore much more sensitive to the sea quark polarization than
$g_1^p(x,Q^2)$ and $g_1^d(x,Q^2)$. It displays a clear double peak
structure, as the sea quarks are dominant in a different $x$-region
than the valence quarks. A precision measurement of $g_1^n(x,Q^2)$
\cite{HERMES} will therefore be able to provide vital new information on
the shape of the sea quark polarization.

 From a consideration of the size of the errors on the
various fitted parameters,
it is apparent that the world data on $g_1(x,Q^2)$  really  only
constrain the polarized valence quark distributions and, to a
lesser extent, the overall magnitude of the
sea quark polarization. The flavour decomposition of the polarized sea
is still completely unknown. Only dedicated experiments, such as the
production of Drell--Yan lepton pairs or the flavour--tagging
of  final--state hadrons in polarized deep inelastic
scattering, will be able to provide further
information. Most important of all,
the polarized gluon distribution is almost completely
undetermined, as its impact on the polarized structure function is
less than the present experimental accuracy.
The variation between our three gluon sets certainly underestimates
the true uncertainty in the distributions.

\section{Prospects for future measurements}

There are several measurements which could be feasible
at future experiments and which yield additional information
on the polarized gluon distribution. In this section we present
some representative predictions for two of these: the
$Q^2$ dependence of $g_1(x,Q^2)$, and an asymmetry  in
$J/\psi$--photoproduction. We also briefly discuss the
prospects of a measurement of
polarized structure functions at the HERA collider.

The derivative of $F_2(x,Q^2)$ with respect to $Q^2$ has been used
to measure the unpolarized gluon distribution in fixed target
experiments \cite{nmc} ($0.008<x<0.5$) and at HERA \cite{heraglu}
($2\cdot 10^{-4}<x<3\cdot 10^{-2}$). The method is particularly
powerful at small $x$, where the gluon distribution dominates the
$Q^2$ evolution,  $\partial F_2 (x)/ \partial \ln Q^2 \sim
P_{qG} (y) \otimes G(x/y)$. In the same way, we can use our three sets of
distributions A, B and C to explore the sensitivity of the polarized
structure function evolution to $\Delta G(x,Q^2)$. Figure~4 shows the
predictions for the asymmetry $A_1(x,Q^2)$ as a function of $Q^2$
in the kinematic range
representative of current fixed--target experiments.
The asymmetry is obtained from Eq.~(\ref{a1def}) with $g_1$ calculated
using our polarized distributions and $F_1$ calculated using the NLO
unpolarized
MRS(A$'$) distributions of Ref.~\cite{mrsg}. The latter are extrapolated
to lower values of $Q^2$, which reproduces the full backwards evolution
to within a few per cent.
At large $x$,
there is no sensitivity to $\Delta G(x,Q^2)$ -- the evolution is completely
dominated by the quark contribution. At small $x$, on the other hand,
we see some dependence  on the gluon.
In this region $\Delta P_{qG} (y) \otimes \Delta G(x/y) < 0$,
and so the derivative
$\partial A_1 / \partial \ln Q^2$ is more negative for the sets that
have a larger gluon polarization above the $x$ value considered.
In particular, we see that at $x \sim 0.01$ the proton asymmetry is almost
$Q^2$ independent for Set C, but decreases with increasing $Q^2$ for
Sets A and B.
Unfortunately, the sensitivity of the present experiments
is much worse than the differences between the various sets. To illustrate
this, we have included two data points at $x = 0.035$ from the
recent SMC and E143 measurements. For the deuteron, the quark contribution
to the structure function is smaller, and so the dependence on $\Delta
G(x,Q^2)$ at small $x$ is somewhat enhanced.
Considering the large errors on the present data, it seems doubtful that
a measurement of the polarized gluon distribution from the
$Q^2$ variation of $A_1(x,Q^2)$ is feasible for values of $Q^2$ where
perturbative expressions can be safely applied. Future high statistics
experiments at higher beam energies \cite{slac15x} will clearly
improve this measurement.
It still has to be kept in mind that a  determination of $\Delta
G(x,Q^2)$ from
the $Q^2$ dependence of $A_1(x,Q^2)$ can never reach the quality of
the corresponding unpolarized measurement, as the gluonic
contribution is not as dominant in the evolution of polarized parton
distributions as it is in the unpolarized evolution \cite{lowx}.

Inelastic $J/\psi$ production from a nucleon target directly
probes the gluon distribution  via the photon--gluon fusion subprocess
$\gamma^{\star}+g\rightarrow(c\bar{c})+g$, which can be described in
the colour--singlet model of Ref.~\cite{colsing}.
Depending on the virtuality of the photon, one can  distinguish two
different classes of events: photoproduction $(Q^2 \approx 0)$ and
leptoproduction $(Q^2 \gg 0)$.
The EMC \cite{EMCjpsi} and NMC experiments \cite{NMCjpsi}
 have obtained measurements of
  the {\it unpolarized} leading--order gluon distribution in the
  range $x \sim 0.05 - 0.25$ from this process.
The results agree well with gluon distributions extracted from other
processes, for example large--$p_T$ direct photon production.
The corresponding cross section for {\it polarized} $J/\psi$ leptoproduction
has been calculated in Ref.~\cite{gui88}.
Taking the photoproduction limit $Q^2 \to 0$, one obtains
\begin{eqnarray}
\frac{\d\sigma_{\lambda h}^{\gamma N} (E_{\gamma})}{\d p_T^2 \d z} &=&
\frac{2\pi}{\alpha}\; \cdot\;  \eta G_{\lambda}(\eta,M_{J/\psi}^2) \;\cdot\;
\frac{8\alpha_s^2
M_{J/\psi} \Gamma_{J/\psi\rightarrow e^+e^-}} {3}\;  \cdot\;
 \frac{z(1-z)}{[M_{J/\psi}^2 (1-z) + p_T^2]^2}  \nonumber \\
 & & \times\  [A(z)+h\lambda C(z)]
\;\cdot \; {\cal F}(p_T^2, z) \; ,
\end{eqnarray}
with
\begin{eqnarray}
{\cal F}(p_T^2, z) & = & \frac{z^2(1-z)^2}{(p_T^2+(1-z)^2
M_{J/\psi}^2)^2}\frac{1}{(p_T^2+M_{J/\psi}^2)^2}\nonumber\\
A(z) & = & \frac{M_{J/\psi}^2}{2}\left(z^2(M_{J/\psi}^2-z s_{\gamma
N})^2 + (1-z)^2(M_{J/\psi}^2+ \right.\nonumber\\
&& \left. (1-z) s_{\gamma N})^2 + (s_{\gamma
N}-M_{J/\psi}^2)^2 \right)\nonumber\\
C(z) & = & (1-z)(M_{J/\psi}^2-z s_{\gamma
N})(zM_{J/\psi}^2(M_{J/\psi}^2-z s_{\gamma N})+(z-2)M_{J/\psi}^2
s_{\gamma N})\nonumber\\
s_{\gamma N} & = & 2 M_N E_{\gamma}\nonumber\\
\eta & = & \frac{p_T^2+(1-z)M_{J/\psi}^2}{z(1-z)s_{\gamma N}}\nonumber\\
z & = & \frac{E_{J/\psi}}{E_{\gamma}}\nonumber\\
\lambda & = & \mbox{gluon helicity}\nonumber\\
h & = & \mbox{lepton helicity}\nonumber
\end{eqnarray}
The polarized (unpolarized) cross sections are obtained by taking the
difference (sum) over the helicity states. The polarized cross section is
then  proportional to $\Delta G (\eta)= G_+ (\eta) - G_-(\eta)$  and
depends only on $C(z)$, while
the unpolarized cross section is  proportional to $G(\eta)=G_+(\eta) +
G_-(\eta)$
and depends only  on $A(z)$.
Taking the  cross sections integrated over one of the variables, one
can define two physics asymmetries:
\begin{eqnarray}\label{eq:Az}
{\cal A}(z) & = & \frac{\d\Delta\sigma^{\gamma N}(E_\gamma)}
{\d z} \left/\frac{\d\sigma^{\gamma N}(E_\gamma)}{\d z}\right.  ,\\
\label{eq:Apt}{\cal A}(p_T^2) & = &  \frac{\d\Delta\sigma^{\gamma N}(E_\gamma)}
{\d p_T^2} \left/\frac{\d\sigma^{\gamma N}(E_\gamma)}{\d p_T^2}\right.  .
\end{eqnarray}
With a photon beam of energy $E_{\gamma}=45$ GeV  and
cuts on $p_T^2$ and $z$ ($p_T^2 > p_{T{\rm min}}^2 = 0.25$~GeV$^2$,
$z < z_{\rm max} = 0.9$, as proposed in \cite{bre94}), one gets
significantly different asymmetries for the three gluon distributions
A, B and C.
Predictions for the asymmetry ${\cal A}(z)$, calculated
using the three leading order distributions in combination with the
unpolarized leading order gluon distributions of Ref.~\cite{grv94},
are shown in Fig.~5.
Note that  small $z$
corresponds to large $\eta$, and the ordering of the asymmetry in
this  region simply reflects the ordering of the different $\Delta G(\eta)$
at large $\eta$. Due to the relatively large asymmetry of up to 15\%,
such a measurement could provide vital information on the leading order
polarized
gluon distribution for $0.1<x<0.35$ and $Q^2\approx M_{J/\psi}^2$.

In the foreseeable future, the HERA collider may be able to
accelerate polarized protons \cite{zeuthen}.
This would offer the unique opportunity of measuring the polarized
structure function $g_1^p(x,Q^2)$ far beyond the $Q^2$ range of present
fixed target experiments. In order to judge the quality of the new
information gained from such a measurement, an order--of--magnitude
estimate of the statistical errors is crucial \cite{bluemlein}.
The structure function $g_1^p(x,Q^2)$ is extracted from a
measurement of the asymmetry
\begin{equation}
A_{\parallel}(x,Q^2) = \lambda_e \lambda_p\;  D(y)\;
\frac{g_1^p(x,Q^2)}{F_1^p(x,Q^2)},
\end{equation}
where $\lambda_e$ ($\lambda_p$) denote the polarizations of the
electron (proton) beam and
\begin{equation}
D= \frac{2y -y^2}{2(1-y)(1+R^p) + y^2}
\end{equation}
is the kinematical depolarization of the photon. The
statistical error on $xg_1(x,Q^2)$ is therefore~\cite{bluemlein}:
\begin{eqnarray}
\delta [ xg_1^p(x,Q^2)] & = &\frac{1}{2\lambda_e \lambda_p}
\frac{1+(1-y)^2}{1-(1-y)^2} \left[2xF_1^p(x,Q^2) +
\frac{2(1-y)}{1+(1-y)^2} F_L^p(x,Q^2)\right] \nonumber\\
& \times & \left( {\cal L}_{int} \int
[ \d^2 \sigma^{\rm (unpol.)}/ (\d x \d
Q^2) ] \d x \d Q^2 \right)^{-1/2} \sqrt{1-A_{\parallel}(x,Q^2)},
\end{eqnarray}
where the unpolarized differential cross section is integrated over
the bin used in the measurement.

To study the accuracy of a measurement of $g_1^{p}(x,Q^2)$ in the collider
mode of HERA, we have evaluated the above expression using the
unpolarized parton distribution set MRS(A$'$) from Ref.~\cite{mrsg} with the
next--to--leading order polarized parton distribution set A described in
Section 2. We apply the following cuts to the HERA phase
space~\cite{klein}: $0.1 < y < 0.95$, $Q^2 > 1\;\mbox{GeV}^2$,
$\Theta_{e'}< 176^{\circ}$ and $E_{e'} > 5\;\mbox{GeV}$, and consider
two scenarios for the beam energies:
\begin{eqnarray}
(a) & : & \sqrt{s} = 300 \;\mbox{GeV} \qquad \mbox{with} \quad
E_e=27.44\;\mbox{GeV},\;E_p=820\;\mbox{GeV},\nonumber \\
(b) & : & \sqrt{s} = 150 \;\mbox{GeV} \qquad \mbox{with} \quad
E_e=18.75\;\mbox{GeV},\;E_p=300\;\mbox{GeV}.
\end{eqnarray}
The expected errors for an integrated luminosity of $60
\;\mbox{pb}^{-1}$ at $\lambda_e=\lambda_p=0.8$ in these two scenarios
are shown in Fig.~6.
We take two bins per decade in $x$ and only one bin in $Q^2$.
These results are consistent with
the leading--order estimates of \cite{bluemlein}, bearing in mind the
different cuts applied. In particular, we estimate smaller errors in
the small--$x$ region, as we apply no cuts on the hadroinc final state.

The average $Q^2$ values probed at HERA range from $2\;\mbox{GeV}^2$
($x\approx 6\cdot 10^{-5}$) to $1060\;\mbox{GeV}^2$ ($x \approx 0.05$) for
$\sqrt{s}=300 \;\mbox{GeV}$ and from $1.7\;\mbox{GeV}^2$ ($x\approx
2 \cdot 10^{-4}$) to $270\;\mbox{GeV}^2$ ($x \approx 0.05$) for
$\sqrt{s}=150 \;\mbox{GeV}$.
For reference, we also show in Fig.~6 a parametrization of $g_1^p(x,Q^2)$
obtained from the NLO set A. To illustrate the impact of a
measurement at HERA, we include the three lowest datapoints
reported by the SMC experiment \cite{exp}, corresponding to
$Q^2$ values around $1.5\;\mbox{GeV}^2$.
It is apparent that a measurement at lower beam energies will yield
data of higher statistical quality. In contrast, the higher beam
energies yield a measurement at smaller $x$.

A measurement of $g_1^p(x,Q^2)$ at the HERA collider will evidently
not provide a large
number of precision data on the $x$ and $Q^2$ dependences of this
structure function. Hence, it will not provide sufficient information
for an indirect determination of the polarized gluon density. The
important physics result of such a measurement is the determination of
the small--$x$ behaviour of $g_1^p(x,Q^2)$, for which various,
significantly different predictions exist \cite{lowxth}. It is
important to stress that a measurement of $g_1^p(x,Q^2)$ at small $x$
will reduce the experimental uncertainty on the Ellis--Jaffe sum rule.
The impact of the small $x$ region can be easily seen in  Fig.~6,
as the Ellis--Jaffe sum rule is proportional to the area enclosed by
$xg_1^p(x,Q^2)$ and the $x$ axis.

A final remarkable point on the measurability of $g_1^p(x,Q^2)$ at
HERA is the impact on the statistical error of the minimum cut on $y$.
 As the photon depolarizes for small $y$, a small cut on $y$
(such as $y>0.01$) diminishes the statistical accuracy,  even though more
data are included in the bin. We find that a minimal cut on $y$
between 0.1 and 0.2 yields the optimal accuracy.

\section{Summary and Conclusions}

We have performed leading and next--to--leading order QCD fits
to the world data on the $g_1$ polarized structure function
measured with proton, neutron and deuteron targets. The quality
of the fits is excellent, and from them we obtain sets of
polarized parton distributions\footnote{The {\tt FORTRAN} code
for the various sets described in this  paper is  available by
electronic mail from T.K.Gehrmann@durham.ac.uk} which can be used
for further phenomenological analyses. The experimental precision
is highest for the proton and deuteron data, and together these
constrain the shapes of the valence $u$ and $d$ distributions.
The sea quark and gluon distributions are still largely undetermined.
There is a weak constraint on the overall size of the former, but almost
no information at all on the flavour decomposition of the sea.
Following our earlier work, we have presented three qualitatively
different gluon distributions, characterized by different behaviours
at large $x$.

It seems that further  progress in the determination of  polarized parton
distributions will require measurements other than $g_1$ at fixed--target
experiments. Inelastic $J/\psi$ photoproduction in
lepton--hadron scattering has already proved capable of measuring
the unpolarized gluon distribution, and a similar measurement with
a polarized beam and target
would provide the first real information on $\Delta G(x,Q^2)$. A measurement of
the $Q^2$ dependence of $A_1(x,Q^2)$ can help to estimate the overall
amount of gluon polarization at higher and medium $x$. Althogh present
experiments cannot provide such a measurement due to low statistics or
too low values of $Q^2$, first information can be expected from the
next  generation of fixed target experiments.
We have also studied the accuracy of a measurement of  $g_1(x,Q^2)$ at
the HERA collider with polarized beams.
Such a measurement will help to discriminate between various
predictions for the behaviour of $g_1(x,Q^2)$ at small $x$, and will
reduce the uncertainty on the Ellis--Jaffe sum rule from the
extrapolation to small $x$.

\section*{Acknowledgements}

\noindent  Financial support from  the UK PPARC (WJS), and from
the Gottlieb Daimler- und Karl Benz-Stiftung and the
Studienstiftung des deutschen Volkes (TG) is gratefully acknowledged.
This work was supported in part by the EU Programme
``Human Capital and Mobility'', Network ``Physics at High Energy
Colliders'', contract CHRX-CT93-0357 (DG 12 COMA). We would like to
thankl Alan Martin, Werner Vogelsang and Dick Roberts for useful
discussions.
\goodbreak

\vfill
\newpage
\subsection*{Figure Captions}

\begin{itemize}

\item[{[1]}] Leading order polarized parton distributions as
described in the text at
$Q_0^2=4\;\mbox{GeV}^2$ compared to the unpolarized distributions of
\protect{\cite{grv94}}.

\item[{[2]}] Next--to--leading order polarized parton distributions
as described in the text at
$Q_0^2=4\;\mbox{GeV}^2$ compared to the unpolarized distributions of
\protect{\cite{mrsg}}.

\item[{[3]}] Structure function measurements $xg_1(x,Q^2)$ of the proton,
deuteron and neutron compared to the next--to--leading order predictions
obtained using Gluon A.

\item[{[4]}] $Q^2$ dependence of $A_1^{p,d}(x,Q^2)$ in next--to--leading
order using the Gluons A, B and C. To illustrate the sensitivity of
current experiments, we  show the $x=0.035$ datapoints from the
recent SMC and E143 measurements.

\item[{[5]}] Predictions for the  asymmetry $A(z)$ (Eq.~(\protect\ref{eq:Az}))
 in polarized $J/\psi$
photoproduction at $E_{\gamma}=45\;\mbox{GeV}$, using the different
leading order polarized gluon distributions.

\item[{[6]}] Expected errors for a measurement of $g_1^p(x,Q^2)$
at the HERA collider with $\sqrt{s}=300\;\mbox{GeV}$ and
$\sqrt{s}=150\;\mbox{GeV}$.

\end{itemize}

\vfill
\newpage

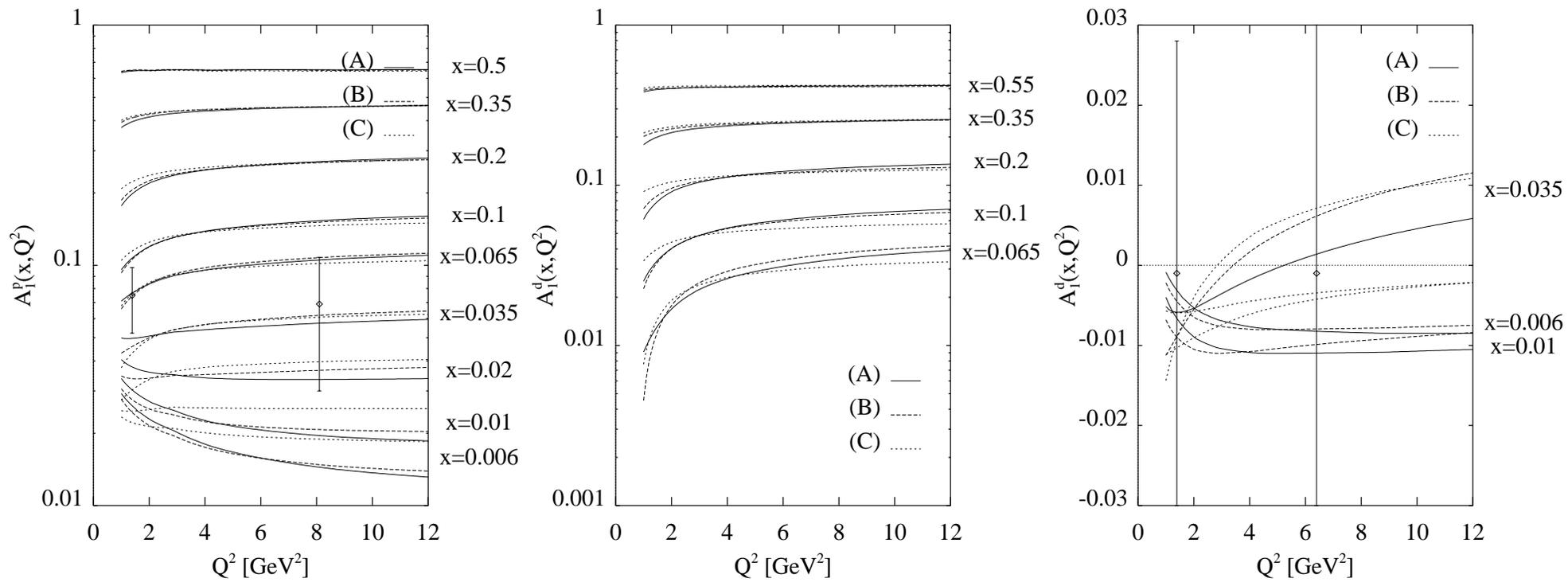
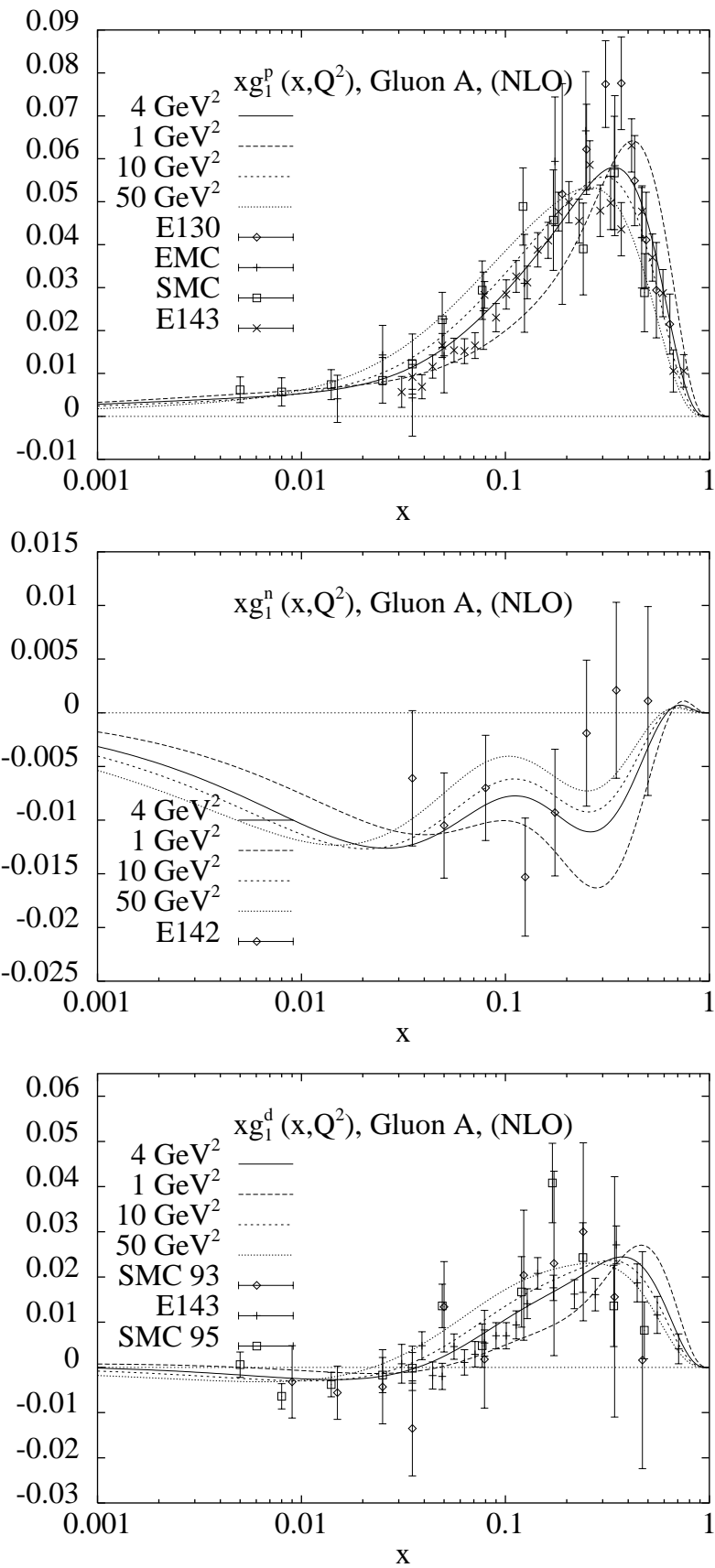
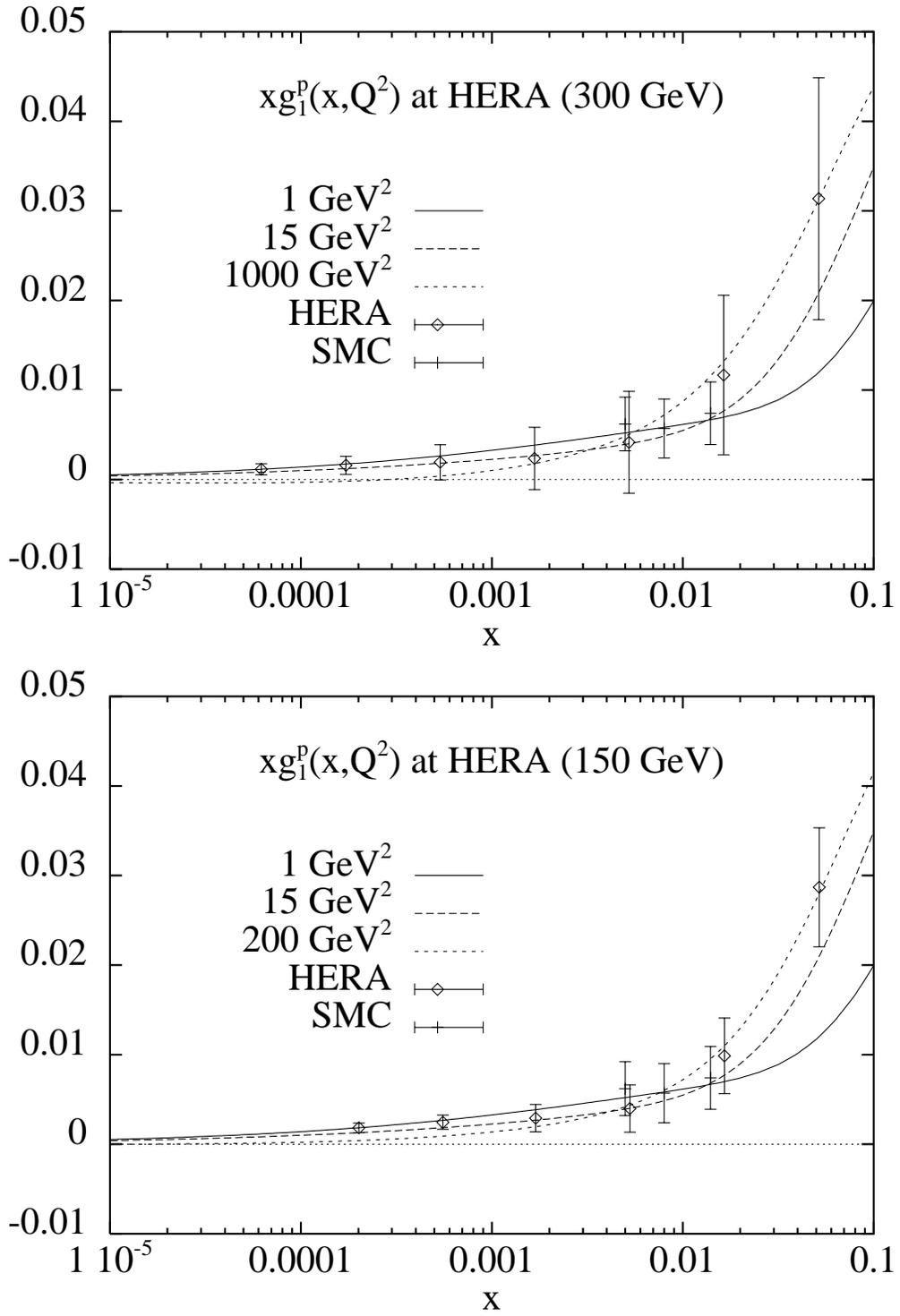
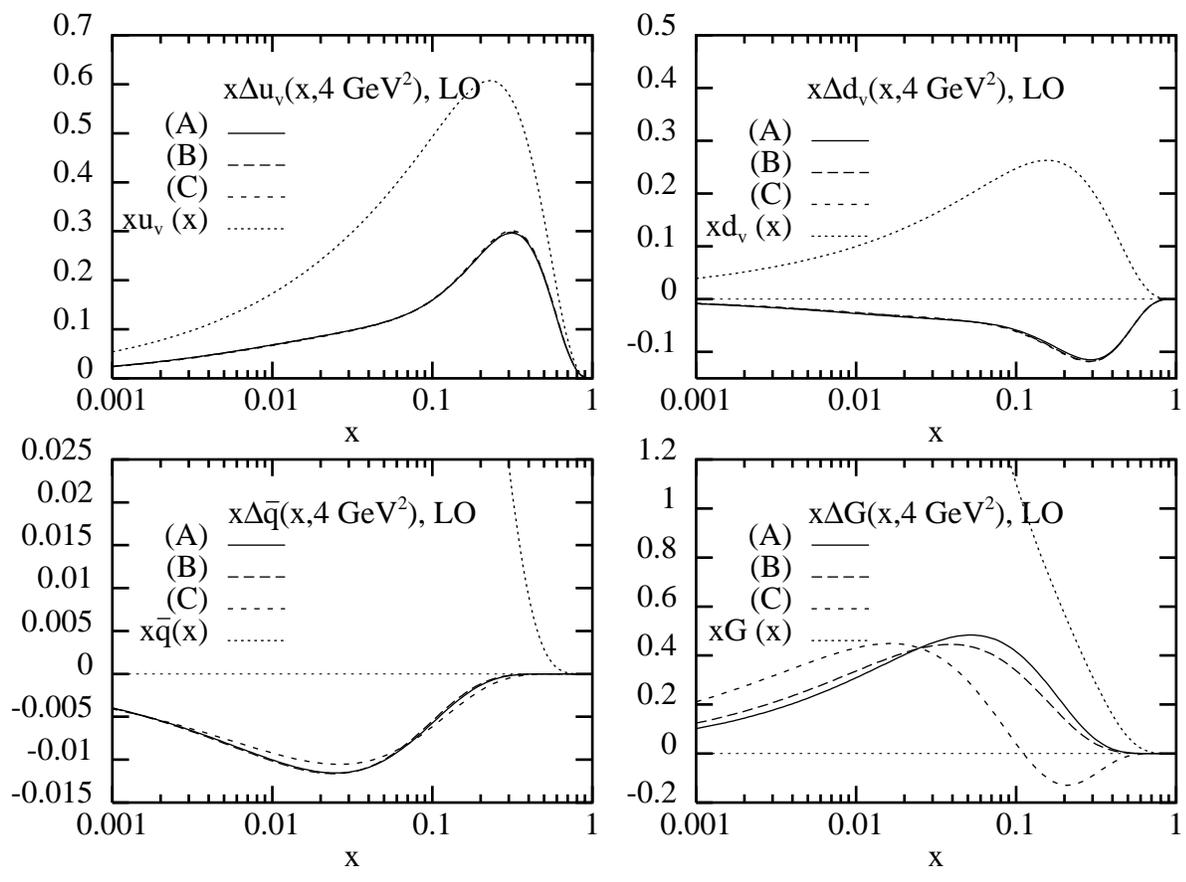
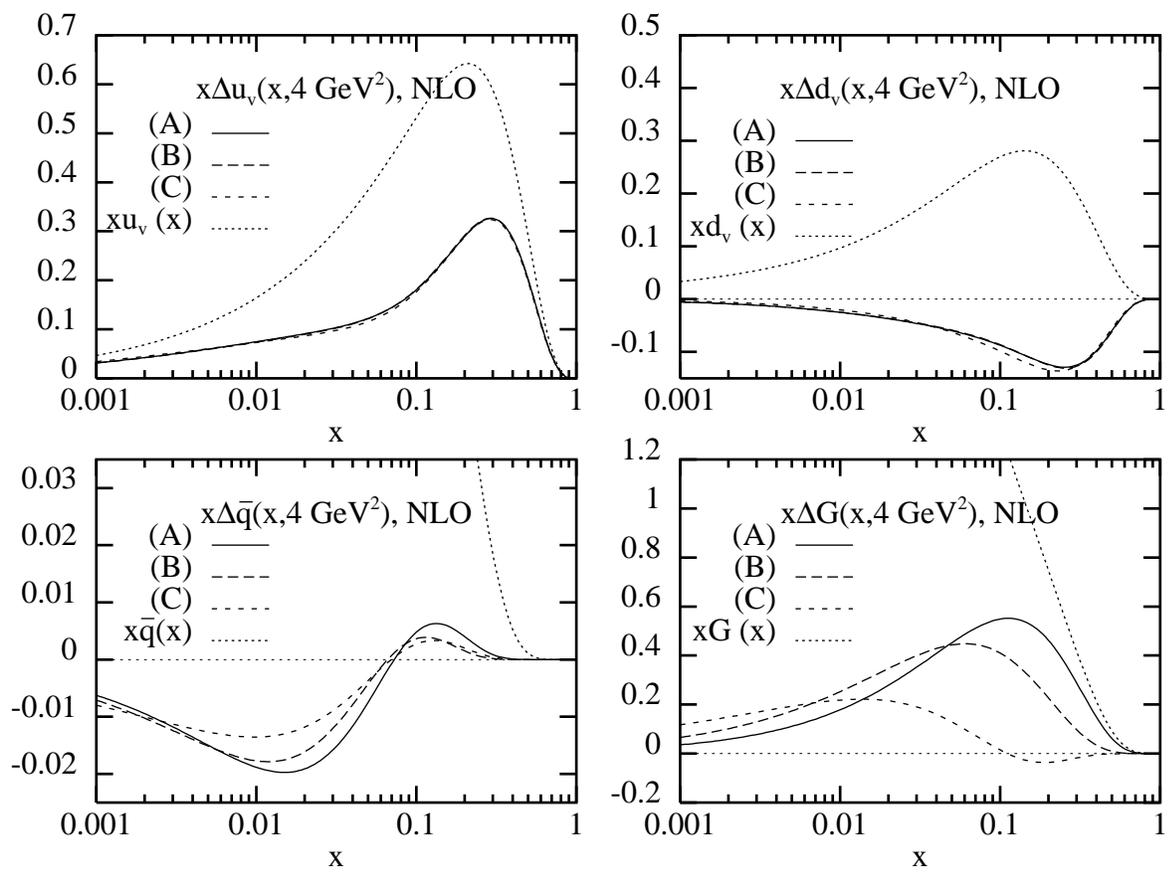
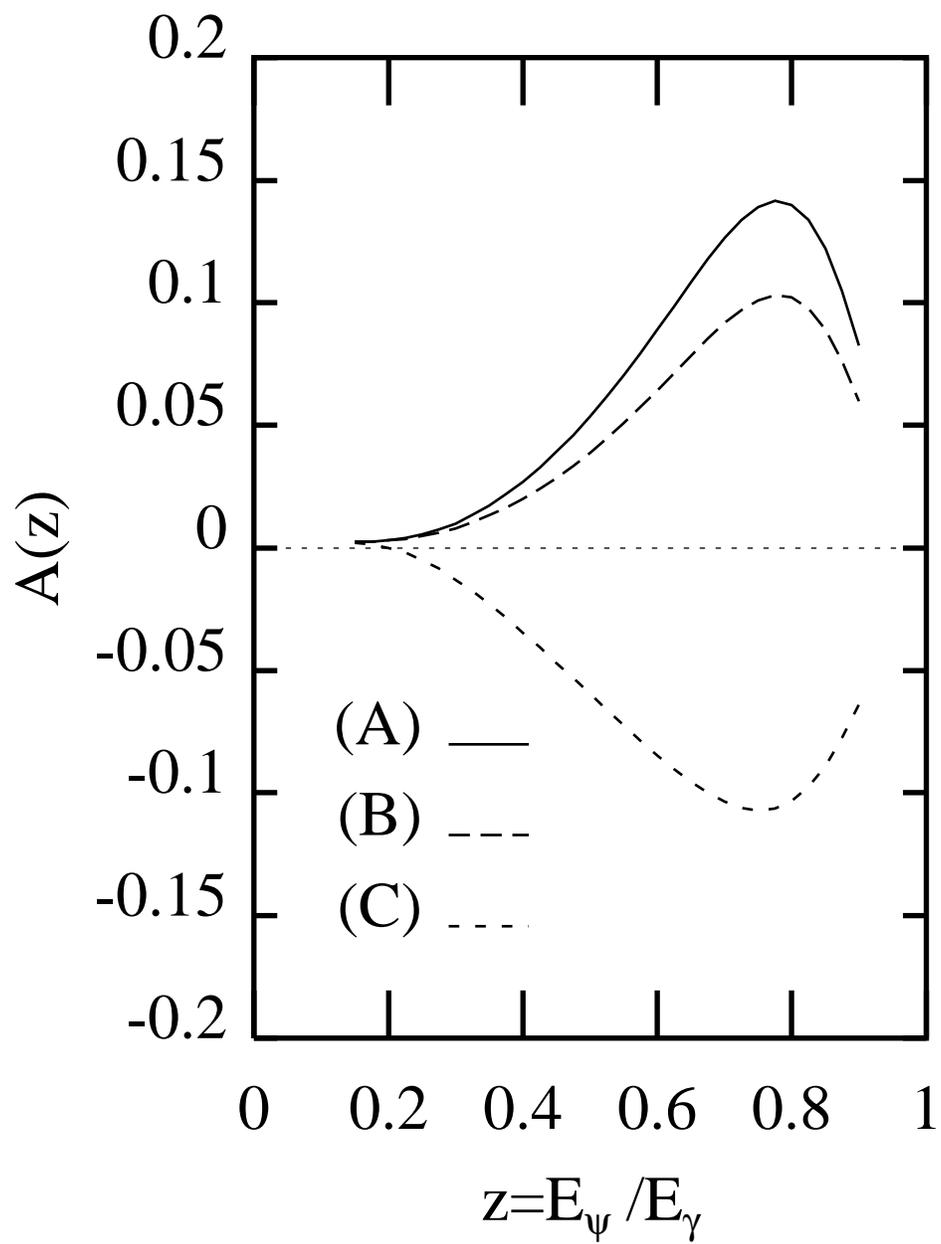
\end{document}